\documentclass[]{nws}

\usepackage{amssymb}
\usepackage{url}

\begin{document}

\title[Face-to-face contacts in an office building]{Data on face-to-face contacts in an office building
suggest a low-cost vaccination strategy based on community linkers}
\author[M. G\'enois, C. L. Vestergaard, J. Fournet, A. Panisson, I. Bonmarin and A. Barrat]{Mathieu G\'enois\email{genois@cpt.univ-mrs.fr}, Christian L. Vestergaard, Julie Fournet\\Aix Marseille Universit\'e, Universit\'e de Toulon, CNRS, CPT, UMR 7332, 13288 Marseille, France;\\
Andr\'e Panisson\\Data Science Laboratory, ISI Foundation, Torino, Italy;\\
Isabelle Bonmarin\\D\'epartement des maladies infectieuses, Institut de veille sanitaire, Saint-Maurice, France;\\
Alain Barrat\\Aix Marseille Universit\'e, Universit\'e de Toulon, CNRS, CPT, UMR 7332, 13288 Marseille, France\\Data Science Laboratory, ISI Foundation, Torino, Italy}

\maketitle

\begin{abstract}
Empirical data on contacts between individuals in social contexts play an
important role in providing information for models describing human behavior and how epidemics spread in populations.
Here, we analyze data on face-to-face contacts collected in an office building.
The statistical properties of contacts are similar to other social situations, but important differences
are observed in the contact network structure. In particular,
the contact network is strongly shaped by the organization of the offices in departments,
which has consequences in the design of accurate agent-based models of epidemic spread.
We consider the contact network as a potential substrate for infectious disease spread  and show that
its sparsity tends to prevent outbreaks of rapidly spreading epidemics. Moreover, we define three typical behaviors according to
the fraction $f$ of links each individual shares outside its own department: residents, wanderers and linkers.
Linkers ($f\sim 50\%$)  act as bridges in the network and have large betweenness centralities. Thus,
a vaccination strategy targeting linkers efficiently prevents large outbreaks.
As such a behavior may be spotted \emph{a priori} in the offices' organization or from surveys, without the full knowledge of
the time-resolved contact network, this result may help the design of efficient, low-cost vaccination or social-distancing strategies.
\end{abstract}

\section{Introduction}

Data-driven models of disease propagation are essential tools for the prediction and prevention of epidemic outbreaks.
Thanks to important increases in data availability and computer power, highly detailed
agent-based models have in particular become widely used to describe epidemic spread at
very different scales, from small communities to a whole country or continent \cite{Davey:2008,Ajelli:2010,Ajelli:2014,Merler:2010}.
One of the interests of  such approaches  comes
from the level of detail they entail in the description of a disease
propagation, from the amount of information they provide on the risk
supported by each population category and on the probabilities of
occurrence of transmission events in different circumstances.
Moreover, the range of modeling possibilities is very large, and many different assumptions can be tested on
how individuals come in contact, how a disease is transmitted and how such
transmission can be contained. The drawback of this freedom in the model design lies in a certain
arbitrariness in modeling choices. In order to alleviate this arbitrariness,
models need to be fed with information and data concerning population statistics and individual
behavior. In this respect, a crucial point regards the way in which people interact in their day-to-day life,
and how these interactions affect disease propagation \cite{Mossong:2008,Read:2012}.
The collection of detailed data sets of human interactions is highly needed, as well as the extraction of
the most relevant stylized characteristics and statistical features of these interactions
\cite{Stehle:2011_BMC,Blower:2011,Machens:2013,Barrat:2013,Barrat:2014}.
Understanding which features of human contact patterns are most salient can also help design
low-cost methods based on limited information for targeted intervention strategies \cite{Lee:2012,Smieszek:2013,Chowell:2013,Gemmetto:2014}.

In order to develop
our knowledge and understanding of human interactions, novel techniques based on sensors using Wi-Fi, Bluetooth or RFID
 have emerged in the last decade and have provided
important new insights  \cite{Zhang:2012,Vu:2010,Eagle:2009,Stopczynski:2014,Cattuto:2010,Barrat:2014,Salathe:2010}.
In the present article, we consider data on face-to-face contacts collected using
wearable sensors  \cite{Cattuto:2010,Barrat:2014}. The corresponding infrastructure, developed by the SocioPatterns collaboration
(www.sociopatterns.org), allows obtaining time-resolved data on close face-to-face proximity events between individuals,
yielding information not only on the overall network formed by these contacts, but also on the
dynamics of these interactions. Previous works have shown that many properties of these dynamics
--contact times, inter-contact times, number of contacts per link, etc -- have broad statistical distributions
and display robust features across several contexts: schools \cite{Salathe:2010,Stehle:2011,Fournet:2014},
hospitals \cite{Isella:2011,Vanhems:2013}, museums \cite{Isella:2011_JTB} or conferences \cite{Isella:2011_JTB,Barrat:2013,Stehle:2011_BMC}.
On the other hand, the contact networks also exhibit different high-level structures in each specific
context, which can influence the way epidemics spread \cite{Isella:2011_JTB,Machens:2013,Gauvin:2013,Gauvin:2014}.
Here we consider the contact patterns between adults at work, which have been less studied in an epidemiological perspective, even if
the influence of office spatial layouts
on social interactions has been considered from a sociological and architectural perspective \cite{Penn:1999,Sailer:2012,Brown:2014,Brown:2014_2}.
\emph{A priori}, the workplace is one of the locations where adults spend most of their time and,
as such, may represent an important spot for transmission of diseases between adults  \cite{Ajelli:2010}. We therefore
present an analysis of the human contact network in such a place, namely a building of the
\emph{Institut de veille sanitaire} (InVS, French Institute for Public Health Surveillance).
We first discuss how the organization of the offices in departments determine the structure of the contact network and the dynamics of the contacts,
and how these features affect epidemic spreading.  We then focus on a new way to determine important
nodes in such a contact network, based on the fraction of links each individual has with other individuals in the
same department and in other departments, and present numerical simulations of an agent-based model of disease spread in order to
discuss the efficiency of vaccination strategies based on this criterion.

\section{Data collection}

In order to collect data on the contacts between individuals, we
used the sensing platform developed by the SocioPatterns\footnote{\protect\url{http://www.sociopatterns.org/}} collaboration, based
on wearable sensors that exchange ultra-low power radio packets in order
to detect close proximity of individuals wearing them \cite{Cattuto:2010,Barrat:2014}.
Each individual that accepted to participate in the study was asked to wear a sensor on his/her chest. As described elsewhere \cite{Cattuto:2010,Barrat:2014}, the body acts as a shield at the radio frequencies used by the sensors, so that
the sensors of two individuals
can only exchange radio packets when the persons are facing each other at close range  ($\lesssim$\,1.5\,m).
Signal detection is set so that any contact that lasts at least $20$ seconds is recorded with a probability higher than $99\,\%$ \cite{Cattuto:2010}.
This defines the time resolution of the setup.

The study took place in one of the two office buildings of the InVS, located in Saint Maurice near Paris, France, and lasted two weeks.
The building hosts three scientific departments -- the \emph{Direction Scientifique et de la Qualit\'e} (DISQ, Scientific Direction),
the \emph{D\'epartement des Maladies Chroniques et des Traumatismes} (DMCT, Department of Chronic Diseases and Traumatisms) and
the \emph{D\'epartement Sant\'e et Environnement} (DSE, Department of Health and Environment) -- along with Human
Resources (SRH) and Logistics (SFLE). DSE and DMCT are the largest departments,
with more than 30 persons each, DISQ and SRH consist of around 15 persons, and finally logistics consists of only 5
persons (Table \ref{tab:number}). DISQ, DMCT and SFLE share the ground floor, while DSE and SRH are located on the first floor.

Two thirds of the total staff agreed to participate to the data collection. The coverage
ranges from $63\,\%$ (DSE, the largest scientific department) to $87\,\%$ (human resources)
(see Table \ref{tab:number}).
A signed informed consent was obtained for each participating individual and the French national bodies responsible for ethics and privacy,
the Commission Nationale de l'Informatique et des
Libert\'es (CNIL, http://www.cnil.fr) was notified of the study.
Data were treated anonymously, and the only information
associated with the unique identifier of each sensor was the department of the individual wearing it.

\begin{table}
  \caption{\label{tab:number}\textbf{Departments of the InVS.}}
  \begin{tabular}{ccccc}
    \hline\hline
         &Number of tags&Number of persons&Coverage&Floor\\
    \hline
    DISQ &15&19&79\%&0\\
    DMCT &30&46&65\%&0\\
    DSE  &38&60&63\%&1\\
    SRH  &13&15&87\%&1\\
    SFLE &4 &5 &80\%&0\\
    \hline
    Total&100&145&69\%&\\
    \hline\hline
  \end{tabular}
\end{table}

\begin{figure}
  \includegraphics[width=\textwidth]{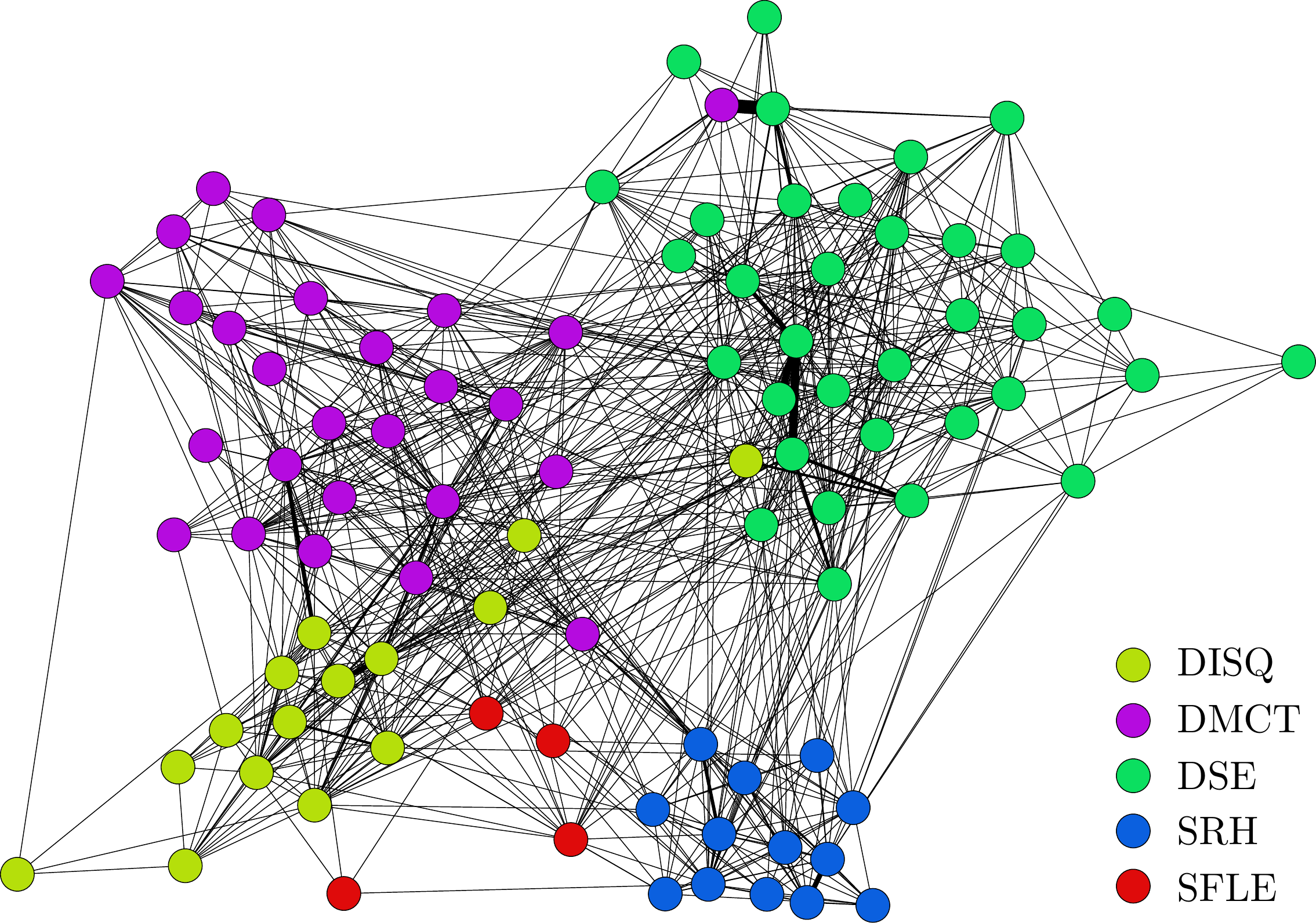}
  \caption{\label{fig:graphe}\textbf{Empirical network of contacts between individuals, aggregated over the two weeks of study.} Each node
  represents an individual and the color-code corresponds to the different departments. Each link between two nodes indicates
  that the corresponding individuals have been in contact at least once during the data collection.
  Nodes are laid out
  using the Force Atlas algorithm (see Gephi software, http://www.gephi.org), which allows
  communities to be apparent in the visualization of the network.}
\end{figure}

\section{Contact dynamics}

\subsection{Aggregated and temporal contact networks}

\begin{figure}
  \includegraphics[width=0.7\textwidth]{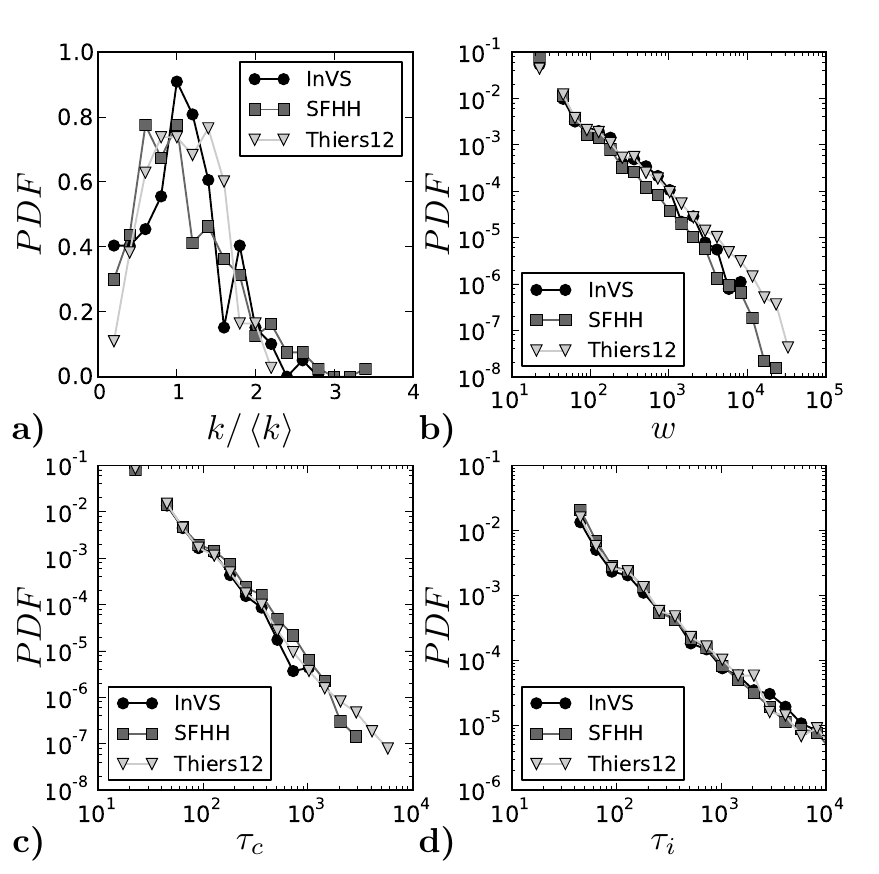}
  \caption{\label{fig:features}\textbf{Main features of the contact data.} For each distribution we compare the features of the present study (InVS) with two previous studies: a scientific conference (SFHH, cf. \protect\cite{Stehle:2011_BMC})
and a high school (Thiers12, cf. \protect\cite{Fournet:2014}).
{\bf a)} Distributions of normalized degrees. Original mean degrees are: 15.1 (InVS), 47.5 (SFHH), 26.4 (Thiers12).
{\bf b)} Link weight distributions. {\bf c)} Contact time distributions.
{\bf d)} Inter-contact time distributions.}
\end{figure}

We build the global contact network, shown in Fig.~\ref{fig:graphe},
by aggregating the contact data over the two weeks of the experiment. Each node represents an individual,
and a link is drawn between two nodes if the corresponding individuals
have been in contact at least once during the study. Each link carries a weight calculated as the total duration of the
contacts between the two individuals. The resulting distributions of node degrees (the degree of an individual gives the
total number of distinct other individuals with whom (s)he has been in contact during the study) and
link weights are shown in Fig.~\ref{fig:features}a\&b.
Moreover, we take advantage of the fact that the data is time-resolved to treat the contact network as a temporal network \cite{Holme:2012}
and compute the distributions of contact durations and of the times between successive contacts of an individual (Fig.~\ref{fig:features}c\&d).

In addition to the statistics of the present data set, Figure \ref{fig:features} also displays the properties of networks of face-to-face contacts
collected in two other settings, namely a conference \cite{Stehle:2011_BMC} and a high-school \cite{Fournet:2014}. Although the contexts  are very different,
the distributions are extremely similar. In particular we find broad distributions with an approximate power-law shape
for weights, contact and inter-contact times, which are typical of the heterogeneous behavior often found in human activities \cite{Barabasi:2005}.

\subsection{Sparsity of the contacts and consequences for the potential spread of infectious diseases}

\begin{figure}
  \includegraphics[width=0.7\textwidth]{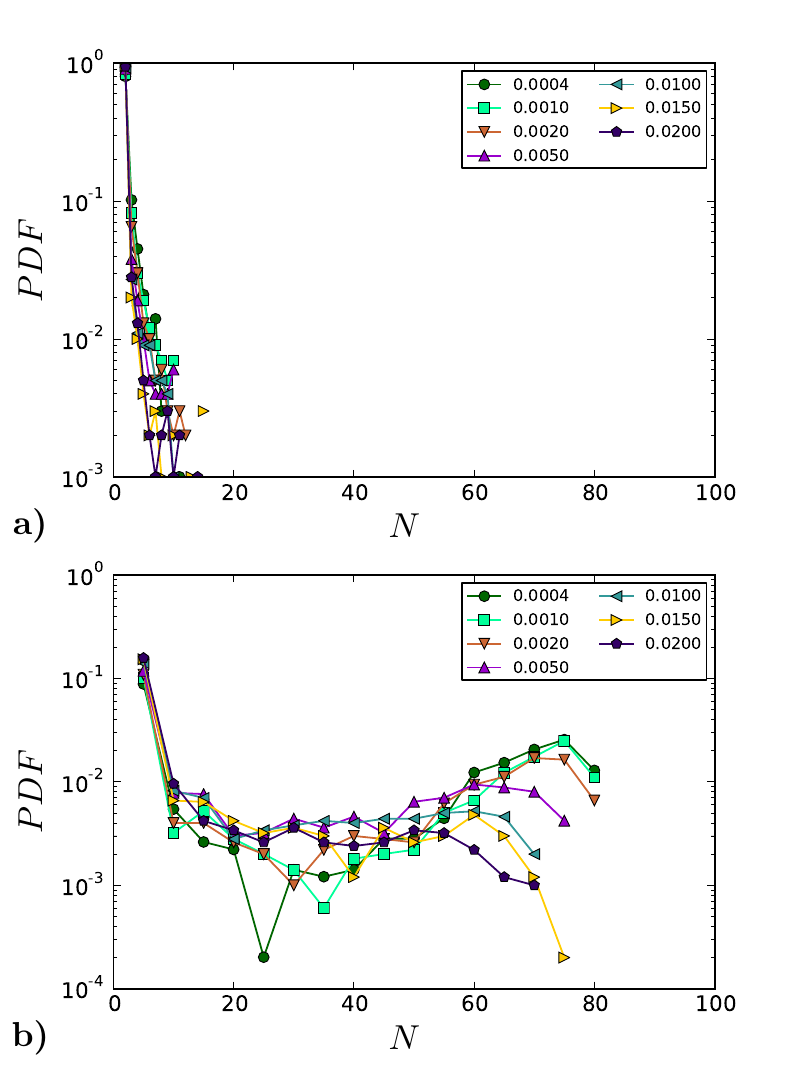}
  \caption{\label{fig:epidemics}{\bf Distributions of the final size $N$ of simulated epidemics.}
  Simulations are performed for different values of the infection rate $\beta$ (given in the legend),
  the recovery rate $\mu$ being determined by the fixed $\beta/\mu$ ratio.  At given $\beta/\mu$, larger values of $\beta$ correspond
  to both faster spread and faster recovery.
  For each value of $\beta$, distributions are computed from $1000$ simulations with initial conditions given by one single randomly chosen infected individual.
  {\bf a)} $\beta/\mu = 100$. {\bf b)} $\beta/\mu = 1000$.}
\end{figure}

One of the main specific features of the present data set is the sparseness of the contacts:
although the study lasted two weeks, the average degree in the aggregated network is only around $15$,
meaning that each individual met only $\sim 15\%$ of the office population participating to the study over the course of these two weeks.
The contact network is thus very far from a fully connected structure, with immediate consequences on possible dynamical processes
taking place on such a network, such as the propagation of an infectious disease.

To explore this issue, we perform numerical simulations of the spread of a Susceptible-Infected-Recovered (SIR) epidemic model in the
population under study. In this model, Susceptible nodes (S) are infected with rate $\beta$ when they are in contact with an infected node (I):
for each small time step $dt$, an S node in contact with $n$ I nodes becomes infected with probability $n\beta dt$.
Infected nodes (I) recover from the infection with rate $\mu$ and enter the Recovered (R) compartment. Recovered nodes cannot be infected again.
Our goal here is not to explore the whole phase diagram of the spreading process but rather to illustrate the influence of the contact patterns
and of the interplay between the time-scales of the contacts and of the spreading process. We therefore consider two different values
of the ratio $\beta/\mu$ and vary the speed of the epidemics by changing $\beta$.  We vary $\beta$ in order to explore a wide
range of time-scales:  $\beta \in [0.0004;0.02]$, which correspond to a typical time of infection $\beta^{-1} \in [50,2500]$ seconds;
For $\beta/\mu = 100$, we thus have $\mu^{-1} \in [1.4;70]$ hours, and for $\beta/\mu = 1000$,
$\mu^{-1} \in [14;700]$ hours. The duration of the spread depends on the value of $\beta$ and ranges from $10^5$ to $10^7$ seconds.

Each simulation starts with a single, randomly chosen, infected node (``seed''). We use the time-resolved data set to recreate \emph{in-silico} the
contacts between individuals (including the periods of inactivity, i.e. nights and week-end, during which individuals
are considered isolated)
and thereby the possibility of the infection to spread. We run each simulation until no infected individual remains
(nodes are thus either still S or have been infected and have then recovered (R)).
We define the duration $T$ of an epidemic as the time needed to reach this state, and the final size $N$ of an epidemic as the number of nodes that
have been affected by the spread, \emph{i.e.} the number of R nodes at the end of the epidemic. As $T$ might be longer than the
duration of the data set, we repeat the two-weeks sequence of contacts in the simulation if needed \cite{Stehle:2011_BMC}.
For each realization we randomly choose both the seed and the moment when the spreading process starts.
We compute the statistics of the final epidemic size over $1000$ realizations for each value of the parameters.

Figure \ref{fig:epidemics} displays the resulting distributions (i.e., the probability that the spread affects $N$ individuals).
For $\beta/\mu = 100$, no large outbreaks are obtained (Fig.~\ref{fig:epidemics}a). Even for the longest
values of the infectious period (lowest values of $\mu$), the sparsity of contacts makes the propagation
of infection difficult. Indeed, at the fastest time scale, \emph{i.e.} the time resolution of $20$ seconds, nodes have an average
instantaneous degree of $0.013$, and overall only $0.66$ links exist on average in the whole network: very few transmission opportunities
exist at each time.
Epidemics spread only if the ratio $\beta/\mu$ is increased enough to compensate for this
very low contact rate (\emph{e.g.}, if, at fixed $\beta$, the recovery rate is strongly decreased).
An example is given in Fig.~\ref{fig:epidemics}b where we use $\beta/\mu = 1000$.
Even in this case, many realizations lead to epidemics of small size, but epidemics affecting a large
fraction of the population are also obtained.

Interestingly, Fig.~\ref{fig:epidemics}b also illustrates the role of the interplay between the timescale of the disease spread and of
the contact network \cite{Isella:2011_JTB,Barrat:2013}. At fixed $\beta/\mu$ indeed,
a mode of the distribution corresponding to large epidemics is present for small values of $\beta$ and $\mu$.
Larger values of $\beta$ and $\mu$
corresponding to faster processes, with high
spreading probability at each contact but also fast recovery and thus shorter infectious periods, lead to smaller probabilities of large epidemics:
as $\beta$ is increased, the mode of the distribution corresponding to large epidemics tends to be suppressed.
This phenomenon is related to the temporal constraints and correlations inherent in temporal networks that can slow down
and hamper the propagation of rapidly spreading processes
\cite{Karsai:2011,Isella:2011_JTB,Pfitzner:2013,Scholtes:2014}: for instance, if $A$ meets $B$ who is infectious, and then $C$,
the infection can spread from $B$ to $C$ through $A$. If instead $A$ meets first $C$ then $B$, $A$ cannot be an intermediary
for the spread to $C$.
Slowly spreading processes --with slow recovery-- on the other hand are  "less constrained" as the infectious period
is longer and contacts with more individuals or repeated contacts with the same individual
can occur during this period, effectively creating more possible paths of transmission between individuals
\cite{Stehle:2011_BMC,Holme:2012,Barrat:2013}.
To make the role of the contact chronology and of temporal constraints more explicit, we compare in the Supplementary Material
the outcome of spreading processes simulated either on the temporally resolved contact data or on aggregated
versions of the data, in which the information about the order in which different contacts occur is lost.
Spreading processes with fixed $\beta/\mu$ values unfolding on a purely static contact structure would
lead to exactly the same distribution of final epidemic sizes for different values of $\beta$.
In our case, we still take into account the fact that nodes are isolated during nights and weekends.  As a result, larger values of
$\beta$  lead to slightly larger epidemics, because faster spreading processes unfold over less inactivity periods
(and inactivity periods, during which nodes can recover but cannot transmit, tend to hinder the spread).
Overall, we thus have an opposite trend on temporal and static networks, at fixed
$\beta/\mu$: on static networks, increasing $\beta$ tends to increase the final epidemic size and, in particular, the
mode at large final sizes is kept, while if the whole temporal resolution is considered,
such an increase tends to suppress the large epidemics.

\subsection{Organization in departments}

The offices are organized in five departments, on two floors. DISQ, DMCT and SFLE (logistics) are on the ground floor,
along with a cafeteria and conference rooms. The remaining departments -- SRH (human resources)
and the DSE department -- are on the first floor. As found in literature linked to social sciences and architecture,
the spatial organization is expected to have an impact on the interactions between office workers
\cite{Penn:1999,Sailer:2012,Brown:2014,Brown:2014_2}. A detailed investigation of this issue is beyond the scope of our work,
in particular as, due to anonymity of the participants, we do not have access to the office location of each participant.
As shown in Fig.~\ref{fig:graphe} the collected data show nonetheless a clear impact of
the structure in departments, while the separation in two different floors is less apparent, at least at this resolution.
As we will see in the next sections and as discussed for instance in \cite{Salathe:2010_PLoS,Hebert:2013}, the fact that departments
seem to correspond to communities in the network structure shown in Fig.~\ref{fig:graphe} has consequences for the possibility
to define mitigation strategies against the spread of epidemics.

\begin{figure}
  \includegraphics[width=\textwidth]{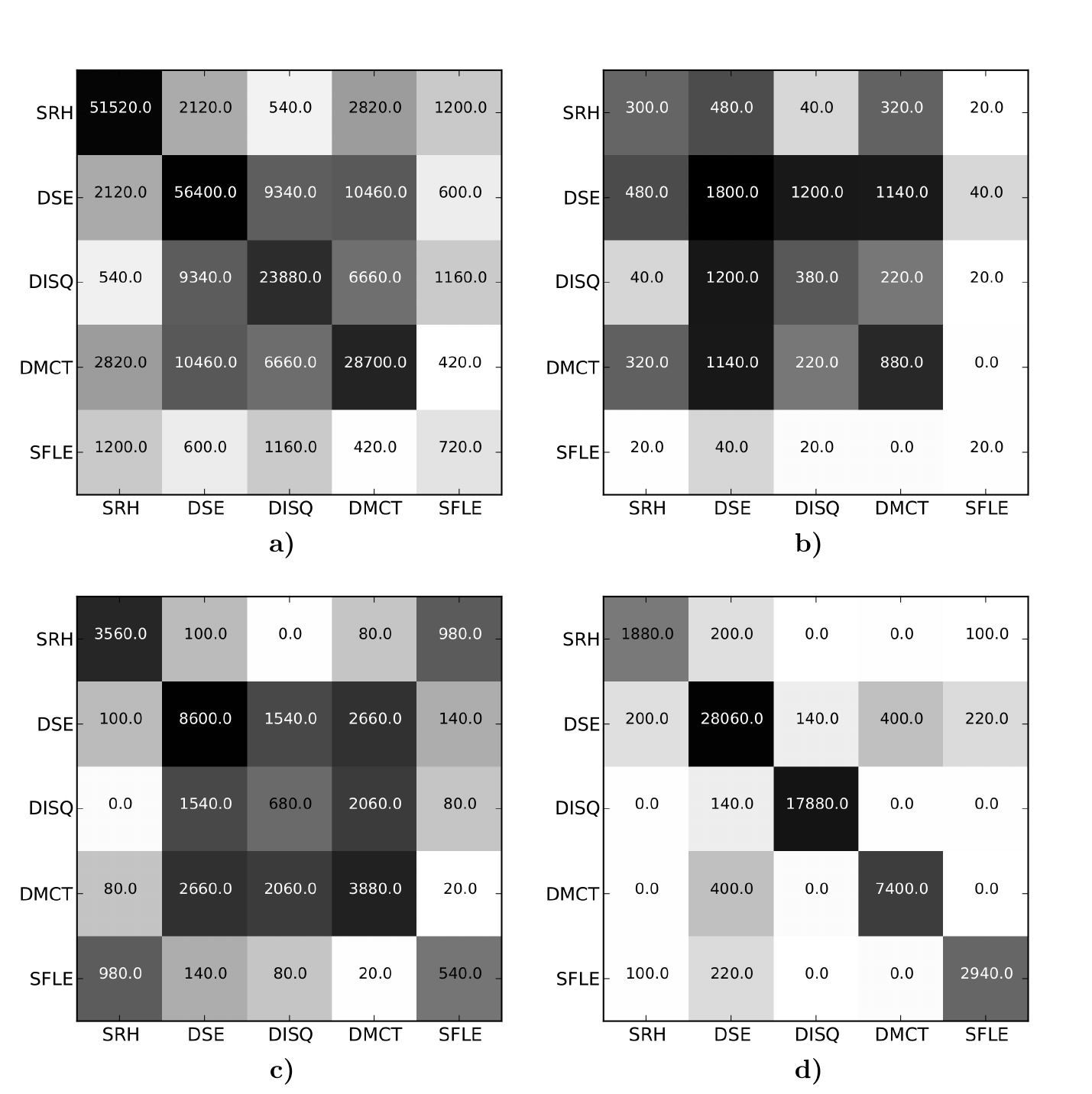}
  \caption{\label{fig:CM}{\bf Contact matrices.} Each matrix element (at row X and column Y) gives
  the total time of contact between individuals from  departments X and Y
  during the two weeks of the study, in different locations.
  {\bf a)} Entire building. {\bf b)} Conference room. {\bf c)} Cafeteria, restricted to the interval between 12am and 2pm for each day.
  {\bf d)} Canteen. This place is in a different building and is not taken into account in a).}
\end{figure}

In order to better quantify the mixing within and between departments, we build
contact matrices giving for each pair of departments the total time of contact between individuals belonging to these departments.
Figure~\ref{fig:CM}a
shows that  contacts occur much more often inside departments (internal contacts) than
between them (external contacts), with the exception of the logistics department.
Furthermore, it confirms that the structure of external contacts does not clearly follow the spatial organization in two floors,
but that scientific departments form a moderately connected sub-structure, leaving human resources and logistics more isolated.
This structure is also found in the contact matrix restricted to the conference rooms (Fig.~\ref{fig:CM}b).
Scientific departments are in fact almost the only ones using these rooms for inter-departments meetings.
Contacts between individuals of different departments also occur independently of scheduled meetings, for instance
during lunch, which takes place either in the cafeteria or in the canteen (located in a distinct building).
The contacts taking place in the cafeteria at lunchtime (Fig.~\ref{fig:CM}c)
show the same structure as the global one, with many internal contacts for each department, and a substructure
corresponding to the three scientific departments on the one hand and logistics and human resources on the other hand.
Strikingly, the structure of contacts occurring in the canteen is rather different, with a mostly diagonal contact matrix:
in this situation, individuals from different departments do not mix with each other. In order to test if such structures of contact patterns can
simply be explained by the fact that the departments are of different sizes or that individuals from the same department tend to be present at a given location at the same time, we consider in the Electronic Supplementary
Material a null model in which contacts occur at random between nodes that are present in the same location at the same time, according to their empirical presence timeline and with a constant rate set to yield the same total contact time as in the empirical data. The empirical contact matrices are significantly different from the ones
obtained with such a null model, showing that the observed structure is indeed the result of individual choices and non-random interactions.

Overall, the structure of contacts measured in the InVS offices is strongly shaped by its internal organization in departments. Although
contacts are \emph{a priori} not constrained by specific schedules, as \emph{e.g.} in schools or high schools \cite{Stehle:2011,Fournet:2014},
the resulting patterns are still very different from contexts such as conferences \cite{Stehle:2011_BMC}, and in particular very far from
a homogeneous mixing situation. This has clear consequences in the design of realistic agent-based models of workplaces. For instance,
as shown in the Electronic Supplementary Material, modeling the spread of a disease in such a context under a homogeneous mixing assumption
would result in a lack of accuracy with respect to more realistic data representations which take into account the division in
departments and the restricted mixing between departments.

\subsection{Weak inter-department ties}

\begin{figure}
  \includegraphics[width=0.7\textwidth]{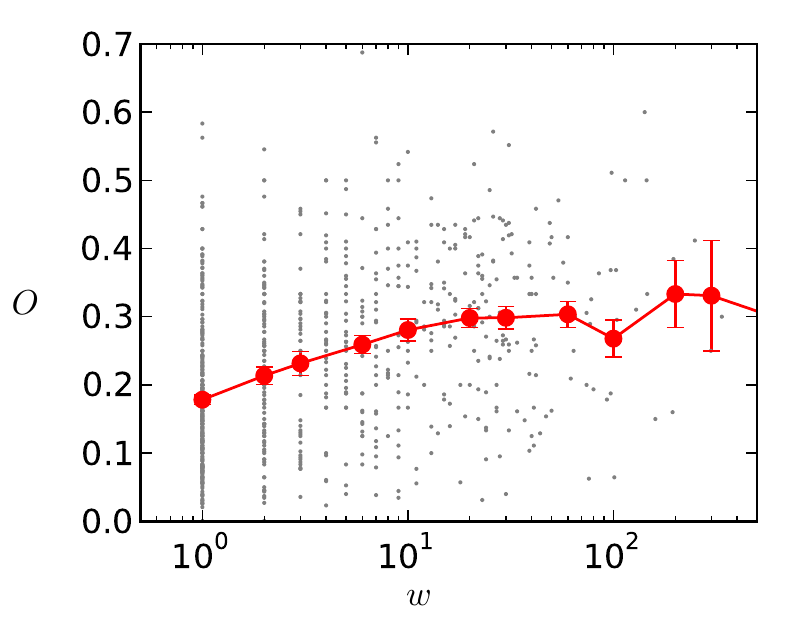}
  \caption{\label{fig:overlap}{\bf Overlap of a link as a function of its weight.}
  Scatter plot (gray) is averaged on weight bins (red). The error bars mark the standard error on the average value.}
\end{figure}

We investigate in more detail how the network structure is shaped by the departments by computing, as
suggested in \cite{Onnela:2007},  the topological overlap $O_{ij}$ of the neighborhoods of each pair of connected nodes $i$ and $j$:
$O_{ij}= n_{ij}/ ((k_i -1) + (k_j-1) - n_{ij})$, where $k_i$ and $k_j$ are the respective degrees of $i$ and $j$ and $n_{ij}$ is the
number of common neighbors of $i$ and $j$.
If $O=0$ the two nodes do not share any neighbors, while if $O=1$ the two nodes have exactly the same neighbors.
The topological overlap thus quantifies whether a link joins two nodes from the same group or community in the network,
or two nodes from different communities.
Figure~\ref{fig:overlap} shows, similarly to  the results of \cite{Onnela:2007} concerning a communication network, that
overlaps and weights are positively correlated (we recall that the weight of a link is the total contact time between the two linked individuals).
Moreover, we measure for  internal links  an average overlap of $0.29\pm?$, with
an average weight of $328\pm?$\,s, while
links joining individuals from different departments have an average overlap of $0.13\pm?$ and an average weight of $134\pm?$\,s.
This indicates that internal links are strong, whereas external links tend to be the weak ties in agreement with the \emph{weak tie hypothesis} of \cite{Granovetter:1973}.

\subsection{Daily contacts structure}

\subsubsection{Stability of the overall structure}

In order to understand how well the structure of the globally aggregated contact network and of the contact matrices
described in the previous paragraph reflect how contacts occur every day, we build
contact matrices giving for each day the durations of contacts between individuals of different departments (shown in the
Electronic Supplementary Material) and compare their properties.
Table~\ref{tab:sim_CM} presents the mean similarity between each daily contact matrix and the matrices of the other days. The high values obtained show that
the structure of the contact matrix is largely conserved from one day to another. As the diagonal of the matrices
contains the largest values, we also compute similarities restricted to the non-diagonal elements. We still obtain rather
large values, showing that even  secondary structures are stable across days.

\begin{table}
  \caption{\label{tab:sim_CM}{\bf Daily contact matrices similarities.}
  For each day, we compute the cosine similarity between the list of elements of the contact matrix and
  the ones of the other days, both with and without the diagonal elements. We list in this table the mean value and the standard deviation of the similarities.}
  \begin{tabular}{ccc}
    \hline\hline
    Day&full&w/o diagonal\\
    \hline
    06/24&0.753 $\pm$ 0.103&0.563 $\pm$ 0.222\\
    06/25&0.843 $\pm$ 0.069&0.481 $\pm$ 0.087\\
    06/26&0.837 $\pm$ 0.046&0.400 $\pm$ 0.246\\
    06/27&0.870 $\pm$ 0.052&0.534 $\pm$ 0.108\\
    06/28&0.871 $\pm$ 0.075&0.426 $\pm$ 0.134\\
    07/01&0.821 $\pm$ 0.058&0.592 $\pm$ 0.152\\
    07/02&0.850 $\pm$ 0.087&0.579 $\pm$ 0.180\\
    07/03&0.858 $\pm$ 0.072&0.488 $\pm$ 0.262\\
    07/04&0.767 $\pm$ 0.058&0.317 $\pm$ 0.131\\
    07/05&0.795 $\pm$ 0.123&0.398 $\pm$ 0.199\\
    \hline\hline
  \end{tabular}
\end{table}

\subsubsection{Network evolution}

\begin{figure}
  \includegraphics[width=0.7\textwidth]{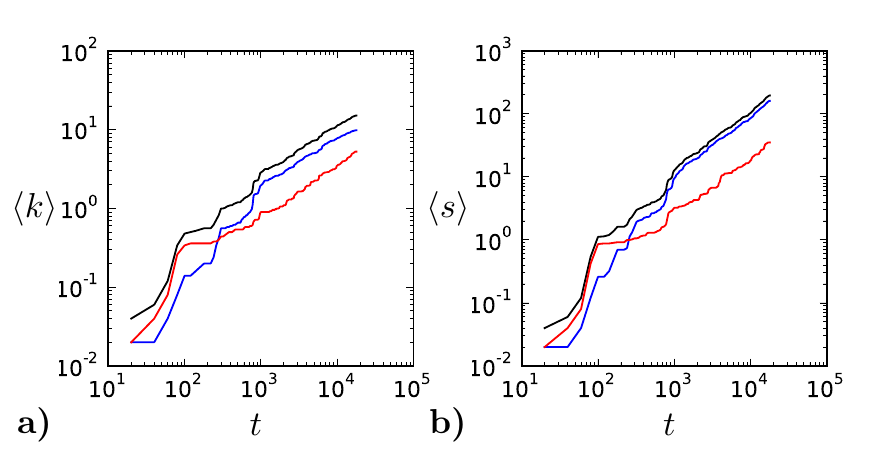}
  \caption{\label{fig:sk}{\bf Network aggregation.} We compute the mean degree (number of distinct individuals contacted) {\bf a)}
  and the mean node strength (total time in contact) {\bf b)} as a function of time while aggregating the network over the two weeks of measurement, disregarding nights and the weekend. These are calculated globally (black), and restricted to internal links (blue) and external links (red).}
\end{figure}

Although the overall contact structure is rather stable from one day to the next,  the specific contacts of each individual
change. Indeed, Fig.~\ref{fig:sk}a shows that the average degree (number of distinct individuals contacted)
of individuals increases steadily when we consider contact networks aggregated over increasing time intervals, both within a department
and for external contacts. In order to gain more
insights into this issue, we
compute similarities between daily aggregated contact networks:
the similarity between two networks $(1)$ and $(2)$ is defined here as the cosine similarity between
their weighted lists of links, $\sum_{i,j} w_{ij,(1)} w_{ij,(2)} / \sqrt{  \sum_{i,j} w_{ij,(1)}^2      \sum_{i,j} w_{ij,(2)}^2     }  $
where $w_{ij,(a)}$ is the weight of the link $ij$ in network $(a)$ (if there is no link, the weight is set equal to $0$).
The resulting values are given in Table \ref{tab:sim}a.

In order to understand if these values correspond to a weak or strong stability, we compare them with several
null models obtained by link rewiring. If we consider networks with the same number of nodes and the same set of weights, but
edges placed at random between nodes (Table \ref{tab:sim}b), similarities are much smaller than for the
empirical data. This is also the case if links are reshuffled while conserving the degree of each node  \cite{Maslov:2002} (Table \ref{tab:sim}c)
and if weights are redistributed at random among the links while conserving the topology of the networks
(Table \ref{tab:sim}d). Finally, even null models which respect the department structure, i.e. with rewiring or reshuffling
procedures inside each compartment of the contact matrix, yield similarities smaller than the empirical ones
(Table \ref{tab:sim}e \& f).

Overall, these results indicate on the one hand that the precise structure of the daily contact networks fluctuate significantly across days, even if the contact
matrix structure is stable. On the other hand, the changes from one day to the next are
 much less important than what would be obtained by random chance, and even some degree of intra-department structure is kept across days.
Such results should thus be taken into account
in the development of agent-based models,  as
done e.g. in \cite{Stehle:2011_BMC}, as the amount of repetition of contacts over different days impacts how dynamical processes
such as epidemics unfold in a population \cite{Smieszek:2009}.

\begin{table}
  \caption{\label{tab:sim}{\bf Daily networks similarities.} For each day, we calculate the cosine similarity
  between the weighted link list of the corresponding network and the lists of all other days. We show here the mean and standard deviation of these similarities. Statistics for the null models are calculated over 1000 realizations. {\bf a)} Empirical similarities. {\bf b)} Complete random rewiring of links. {\bf c)} Rewiring keeping each individual node degree fixed. {\bf d)} Random redistribution of link weights on the fixed topological structure.
  {\bf e)}  Rewiring in each contact matrix compartment. {\bf f)} Random redistribution of link weights in each contact matrix compartment.}
  \begin{tabular}{ccccccc}
    \hline\hline
    &{\bf a)}&{\bf b)}&{\bf c)}&{\bf d)}&{\bf e)}&{\bf f)}\\
    Day&& ($\times 10^{-3}$)&($\times 10^{-2}$)&($\times 10^{-2}$)&($\times 10^{-2}$)&($\times 10^{-2}$)\\
    \hline
    06/24&0.141 $\pm$ 0.084&6.20 $\pm$ 13.8&1.51 $\pm$ 1.56&3.83 $\pm$ 3.50&2.19 $\pm$ 2.69&4.39 $\pm$ 3.63\\
    06/25&0.302 $\pm$ 0.162&5.30 $\pm$ 13.5&2.40 $\pm$ 4.46&4.07 $\pm$ 3.90&3.56 $\pm$ 5.41&6.34 $\pm$ 6.59\\
    06/26&0.125 $\pm$ 0.052&5.33 $\pm$ 15.6&2.29 $\pm$ 3.44&3.96 $\pm$ 3.99&2.90 $\pm$ 3.75&5.15 $\pm$ 4.78\\
    06/27&0.288 $\pm$ 0.195&6.32 $\pm$ 14.8&2.01 $\pm$ 2.09&4.67 $\pm$ 3.65&3.48 $\pm$ 4.73&6.27 $\pm$ 6.10\\
    06/28&0.096 $\pm$ 0.041&5.30 $\pm$ 12.8&2.23 $\pm$ 3.43&4.98 $\pm$ 4.08&4.01 $\pm$ 5.38&7.55 $\pm$ 6.55\\
    07/01&0.109 $\pm$ 0.078&4.15 $\pm$ 11.9&1.67 $\pm$ 2.93&4.07 $\pm$ 4.08&2.19 $\pm$ 2.35&4.75 $\pm$ 3.65\\
    07/02&0.264 $\pm$ 0.141&5.85 $\pm$ 14.7&2.65 $\pm$ 3.92&5.28 $\pm$ 4.10&4.36 $\pm$ 4.76&8.06 $\pm$ 5.83\\
    07/03&0.318 $\pm$ 0.206&5.23 $\pm$ 14.3&2.03 $\pm$ 2.35&4.56 $\pm$ 4.04&3.36 $\pm$ 5.22&6.72 $\pm$ 7.07\\
    07/04&0.246 $\pm$ 0.101&5.03 $\pm$ 13.9&1.62 $\pm$ 2.19&3.94 $\pm$ 3.90&2.46 $\pm$ 3.22&5.23 $\pm$ 4.64\\
    07/04&0.297 $\pm$ 0.262&2.52 $\pm$ 11.4&1.37 $\pm$ 4.11&2.43 $\pm$ 3.96&2.75 $\pm$ 6.37&4.79 $\pm$ 8.05\\
    \hline\hline
  \end{tabular}
\end{table}

\subsubsection{Daily activity}

\begin{figure}
  \includegraphics[width=0.8\textwidth]{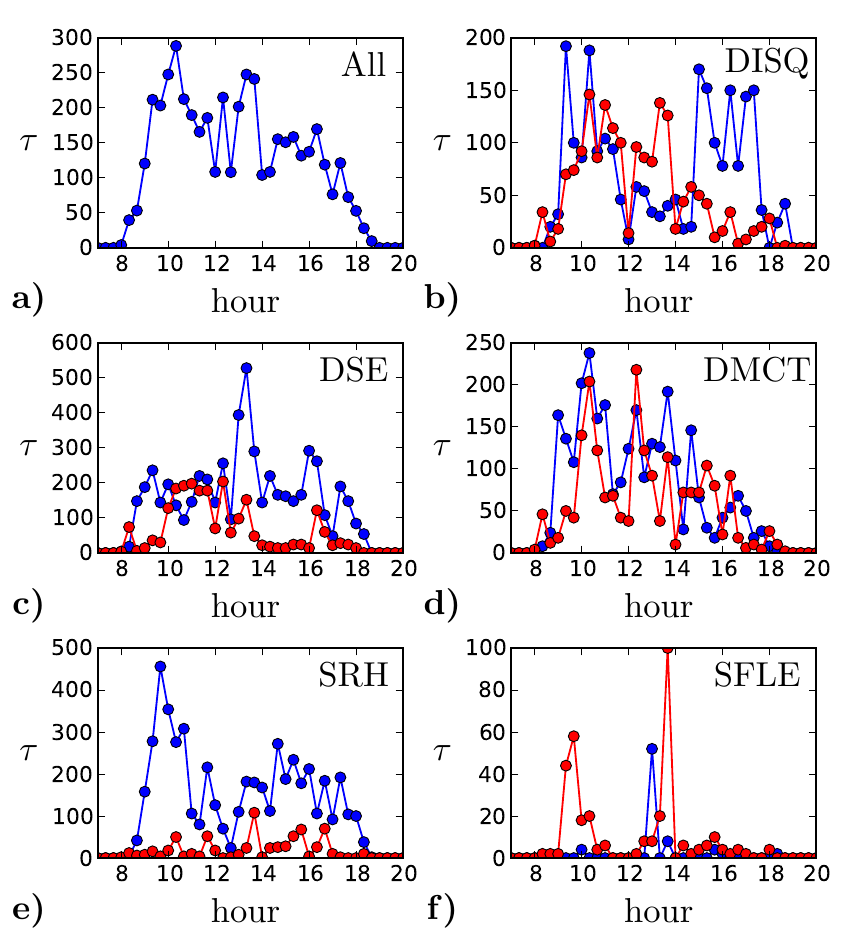}
  \caption{\label{fig:activity}{\bf Daily activity timelines for each department.} The activity is defined as the total duration of
  contacts during windows of 20 minutes. We show the activity averaged over the ten days of the study. Internal contacts are plotted in blue, external contacts in red.}
\end{figure}

As Fig.~\ref{fig:sk}b shows, the total duration of contacts within and between departments grows steadily. Contacts however might
\emph{a priori} occur at specific times for specific departments (due to breaks, meetings, etc...). We therefore examine in Fig.~\ref{fig:activity}
the activity timelines averaged over different days. The global, averaged activity for the whole building (Fig.~\ref{fig:activity}a) exhibits activity
peaks around 10am and lunchtime, but no other clear feature emerges. This is most probably due to the fact that no strict schedule
constraints exist (in contrast \emph{e.g.} to schools \cite{Stehle:2011} and hospitals \cite{Vanhems:2013}). We also show for
 each department both internal and external contact activities for an average day (Fig.~\ref{fig:activity}b to f).
 Four departments present the 10am peak of activity, which may be related to a common
 tendency to have a break around that time. During lunch (12am-2pm), DSE, DMCT and SFLE have many contacts,
 both within and outside the department, whereas SRH exhibits a decrease of activity, and
 DISQ shows an inversion between internal and external contacts.
 Finally, for the three scientific departments, external contacts are more present in the morning than in the afternoon.

\section{Node behaviors}

\subsection{Residents, wanderers and linkers}

\begin{figure}
  \includegraphics[width=\textwidth]{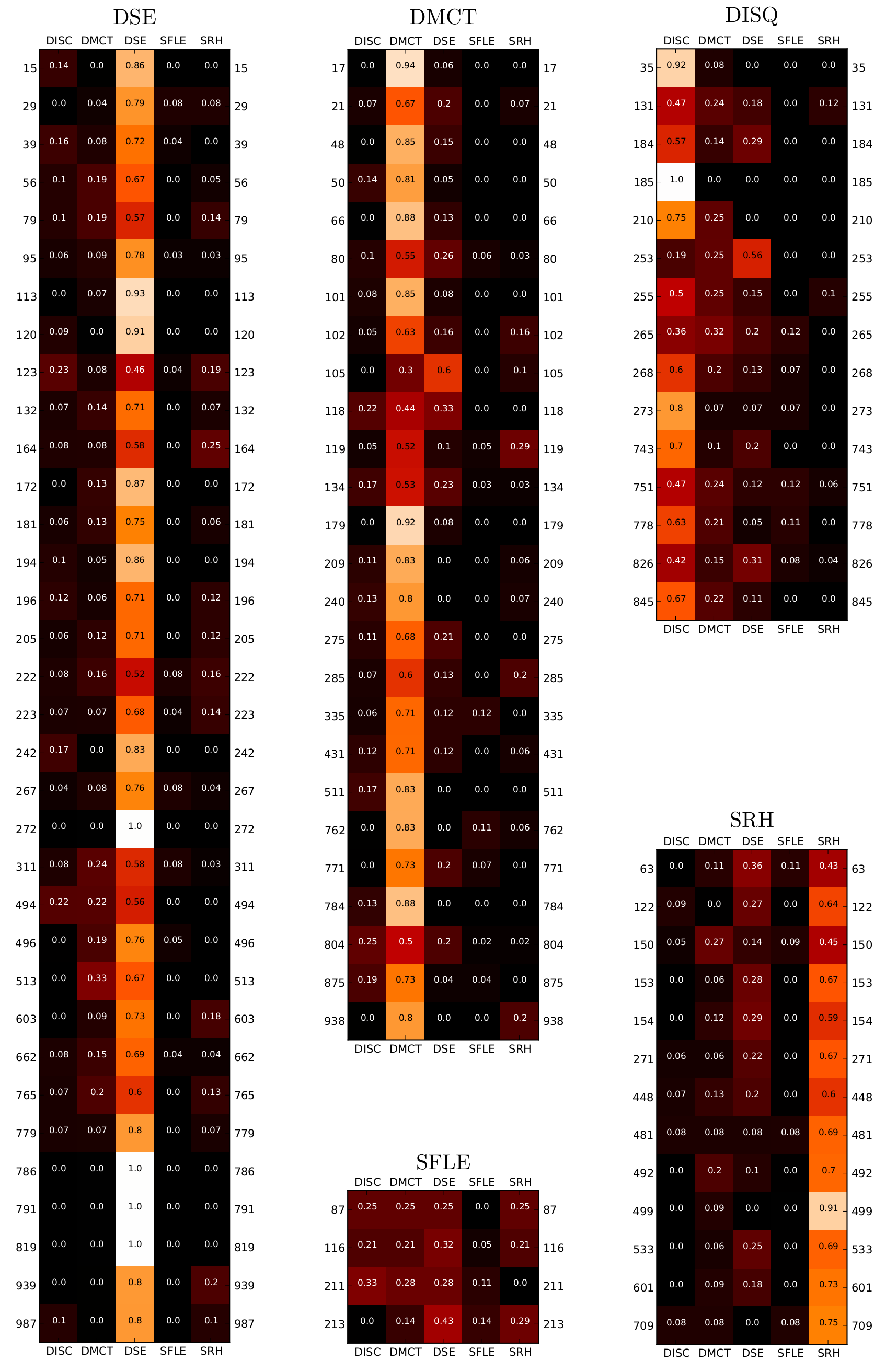}
  \caption{\label{fig:Prop}{\bf Proportion of links with individuals of each department, for each node.}
  Values are calculated from the network aggregated over the two weeks.}
\end{figure}

In the context of spreading phenomena, links joining groups of individuals in a contact network
play a crucial role \cite{Karsai:2011,Onnela:2007,Salathe:2010_PLoS,Hebert:2013}. In the present case,
the possibility for an infectious disease to spread in the whole population thus
strongly depends on the structure of contacts along inter-departments links.
In order to shed more light on the role of each node in linking communities, we consider the static aggregated network, and we compute in Fig.~\ref{fig:Prop}
for each node the fraction of its links with individuals of each department. We focus in particular on the fraction $f$ of internal links
(i.e., with nodes in the same department). Most nodes are ``residents'': the large majority of their links are internal.
Some other nodes (such as
individuals from the Logistics department, or node 105 from DMCT and node 253 from DISQ) are on the other hand mostly linked
to nodes outside their own department: they are ``wanderers''.

\begin{figure}
  \includegraphics[width=0.7\textwidth]{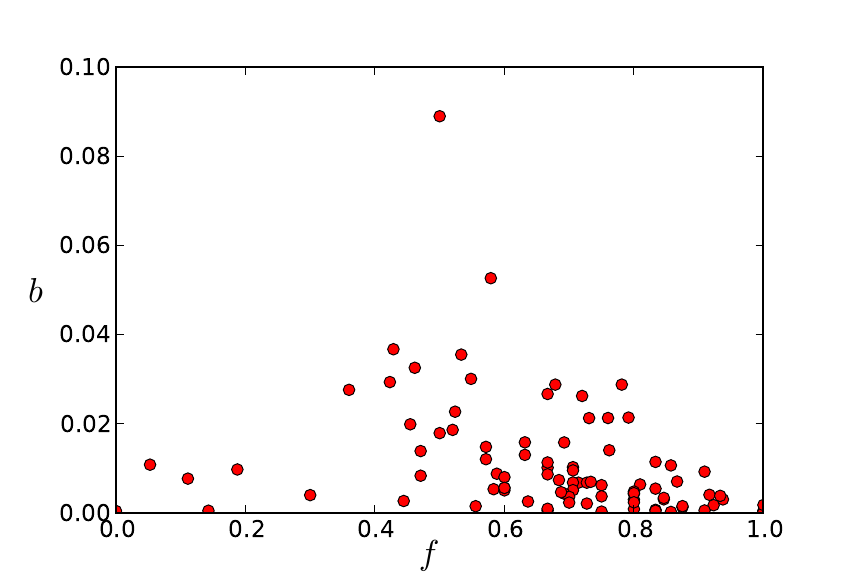}
  \caption{\label{fig:p_ratio}{\bf Relation between the betweenness centrality $b$ of
  a node and the fraction $f$ of its links that are internal.} Values are calculated on the network aggregated over the two weeks.}
\end{figure}

In order to determine how these different types of behaviour could be exploited in the design of targeted vaccination strategies, we investigate
in Fig.~\ref{fig:p_ratio}  the relation between the centrality of a node in the aggregated contact network,
as measured by its betweenness centrality (BC), and its fraction $f$ of internal links.
We note here that different definitions of node centrality exist. They are often correlated but the precise ranking of nodes' centrality slightly
depends on the specific  centrality measure used, and various works have compared how different centrality measures perform in identifying
individuals most at risk of infection or the most efficient spreaders for various models of epidemic spreading
\cite{Christley:2005,Kitsak:2010,Castellano:2012}. In the context of targeted measures to mitigate the spread of diseases in particular,
immunizing (i.e., removing) the nodes with highest degree or largest
BC is known to be among the most efficient strategies \cite{Pastor-Satorras:2002,Holme:2002,Dall'Asta:2006}. We also note that BC can
be defined both on the unweighted and weighted versions of the network  \cite{Dall'Asta:2006,Opsahl:2010}. We have considered both
cases and found very similar results so we use for simplicity in
Fig.~\ref{fig:p_ratio} the unweighted BC.

Residents have, as could be expected, low centrality.
Wanderers, as nodes connecting to other departments, could be expected to be crucial in potential spreading paths. However,
they  turn out to have low centralities as well.
The most central nodes correspond to a specific population, composed of nodes whose neighborhood is composed
by approximately one half of internal links and one half of external links. We call them ``linkers'', as they are effectively responsible for the
connectivity between the departments and act as bridges \cite{Wasserman:1994} between them.

\subsection{Linkers and spreading processes}

\begin{figure}
  \includegraphics[width=\textwidth]{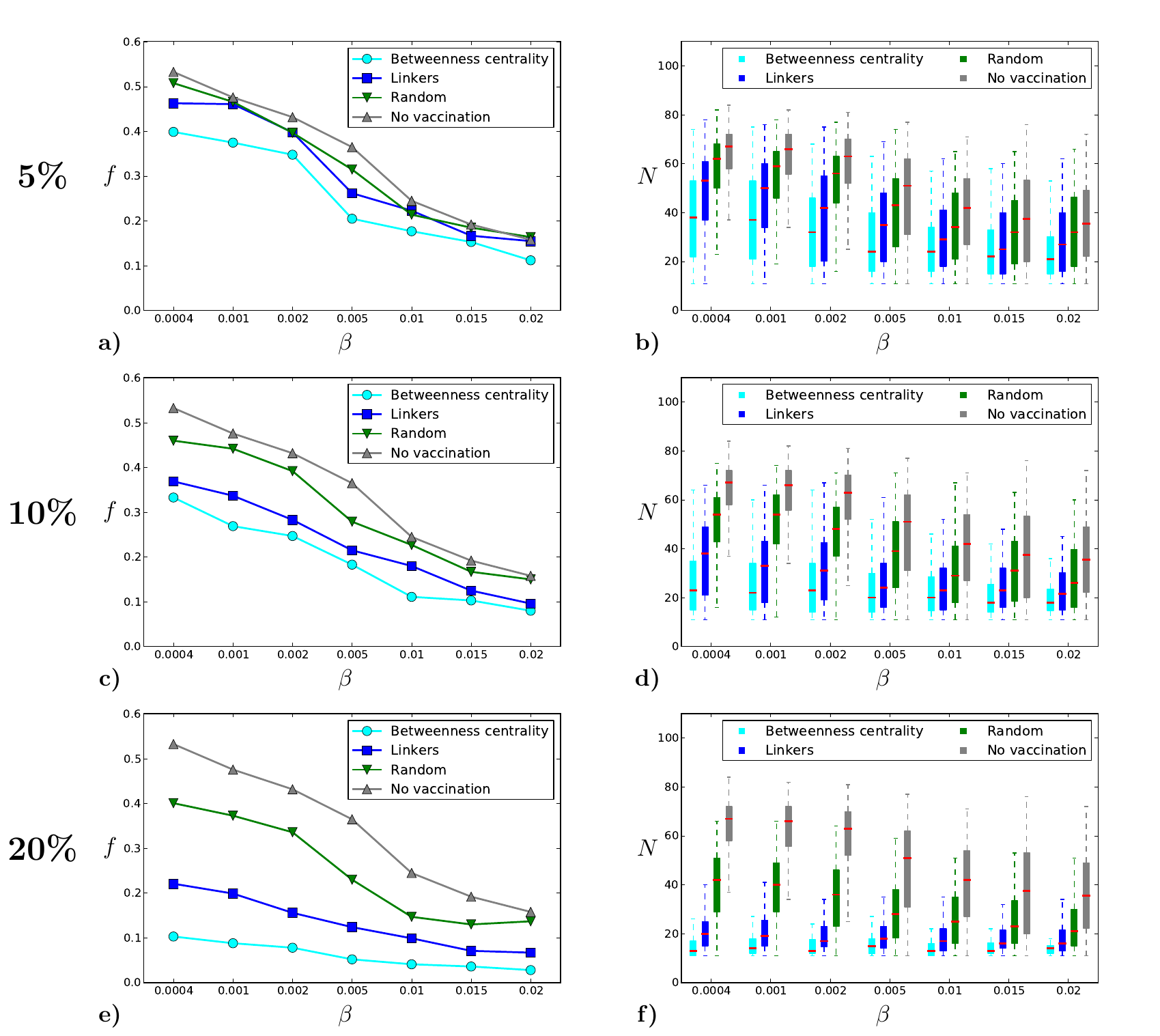}
  \caption{\label{fig:vacc}{\bf Vaccination strategies.} We simulate SIR epidemics on the time-resolved contact network, with $\beta/\mu = 1000$,
  in order to test different vaccination strategies. In each case we perform $1000$ realizations. {\bf a)}, {\bf c)}, {\bf e)} Fraction $f$ of epidemics
  that reach more than 10 individuals. {\bf b)}, {\bf d)}, {\bf f)} Size distribution of these large epidemics.
  For each strategy we vaccinate nodes according to the following rules: highest betweenness
  centralities, best linkers (fraction of internal links closest to $0.5$) and random choice. We also plot the case without vaccination as a reference.}
\end{figure}

As the linker behavior seems to be associated with high node centralities in the aggregated contact network, we consider here the effect on a potential
epidemic spread of a containment strategy targeting linkers. Although it is expected that the precise detection (and vaccination) of the nodes with
the highest centrality would be more efficient, as the correlation between linkers and high centrality nodes is not perfect, the detection of linker
behavior relies \emph{a priori} on less information than the computation of betweenness centralities and might thus represent an interesting alternative.
We show in Fig.~\ref{fig:vacc} the result of simulated SIR processes on the temporal contact network when a fraction of nodes are vaccinated, i.e. considered
immune to infection (they can neither be infected nor transmit the infection), according to different strategies. The figure shows
that the targeted vaccination of linkers decreases the probability to observe a large outbreak, and, in the
case of such an outbreak, limits its size. This strategy is much more efficient than random vaccination, and,
for a vaccination rate of  $20\,\%$, performs almost as well as a strategy based on centrality, with a strongly decreased outbreak probability.
For a small vaccination rate ($5\%$), the outbreak probability is not much reduced, but, in case of outbreak, the resulting epidemic sizes are
clearly smaller than for a random vaccination strategy.

\subsection{Stability of the linker behavior}

\begin{figure}
  \includegraphics[width=0.7\textwidth]{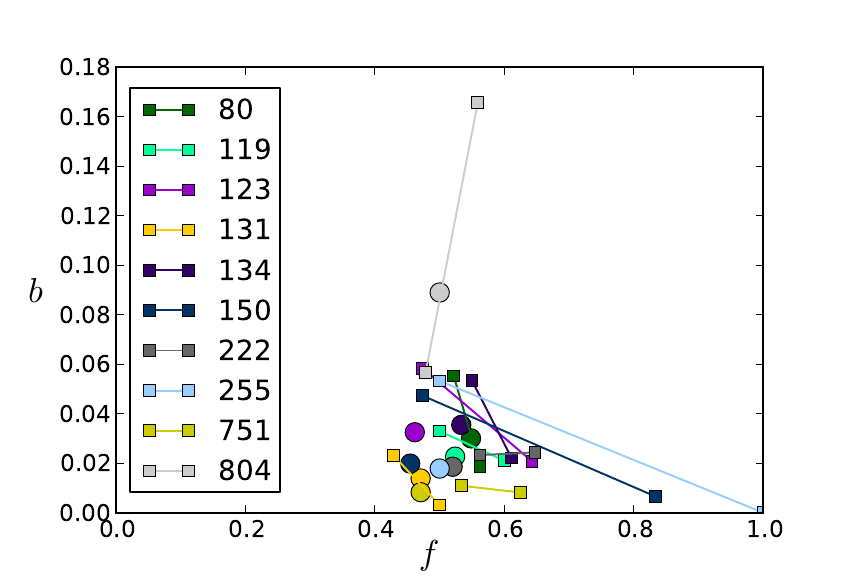}
  \caption{\label{fig:traj}{\bf Stability of linker behavior during the two weeks.} For each node defined
  as a linker from the two weeks of data, we plot its characteristics based on each week's data (squares).
  Circles show their positions according to the full two weeks of data.}
\end{figure}

\begin{figure}
  \includegraphics[width=\textwidth]{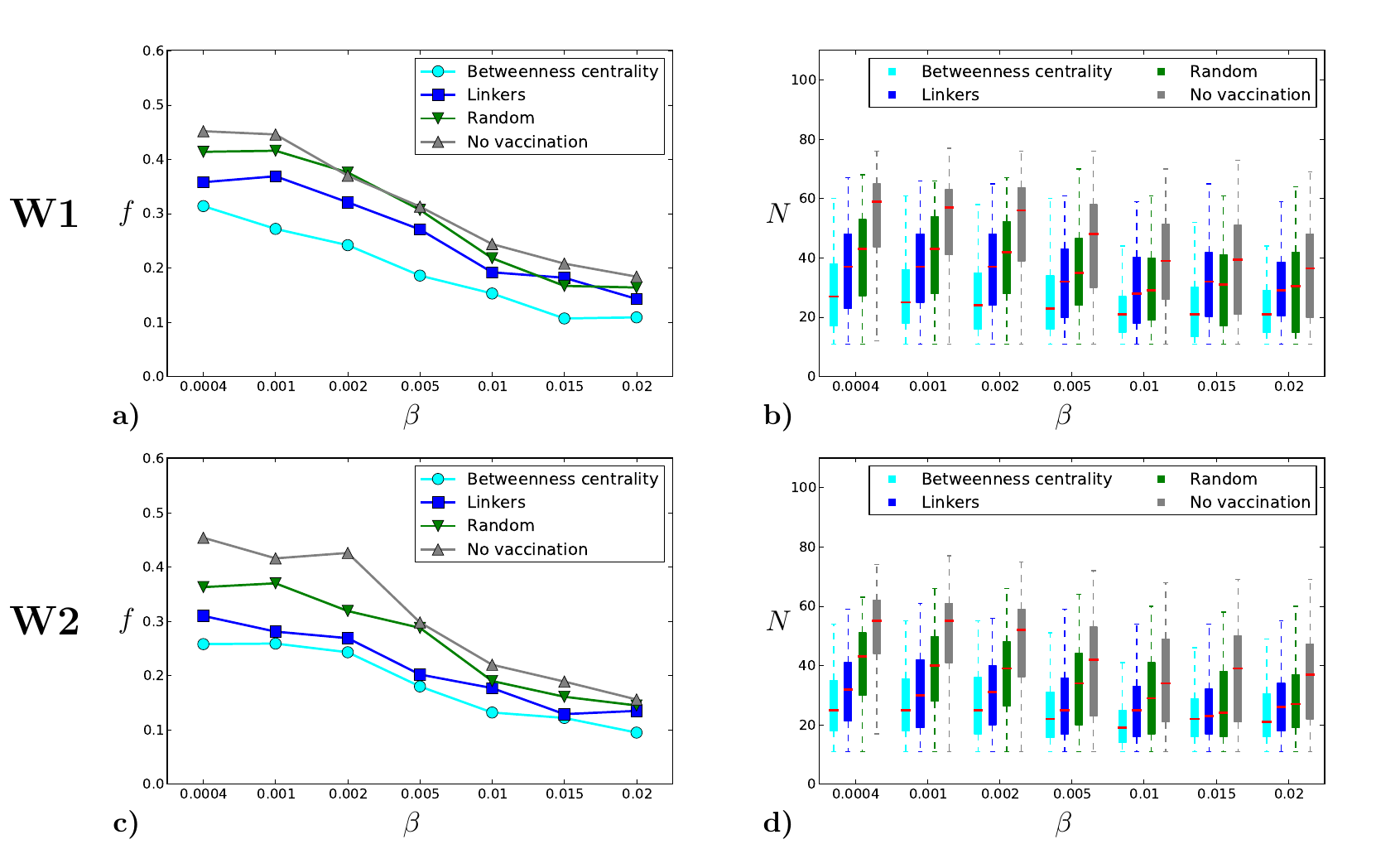}
  \caption{\label{fig:vacc_S}{\bf Efficiency of sampling over one week.} We simulate SIR epidemics on the  temporal
  contact network, with $\beta/\mu = 1000$, over only one week of data, with vaccination strategies based on data from the other week. In each case 10\,\% of the population was vaccinated. The labels indicate on which week the vaccination strategies were defined.}
\end{figure}

Linkers are defined as nodes with approximately $50\,\%$  external links, based on the network aggregated over the two weeks data set.
If we perform the same analysis on data restricted to either of the weeks, we do not necessarily find the same nodes, as the fine structure of the networks fluctuates a
lot from one day to another, as highlighted in the previous section. We investigate this point in Fig.~\ref{fig:traj} by plotting, for each node
with a linker behavior in the global two weeks aggregated networks, its betweenness centrality versus its fraction of external links in each of the two
networks aggregated over one week.
Though some of these nodes would not be selected as linkers if we restricted the data to one week,
their behavior can still be considered as linker-like, as the fraction of external links  remains between $40\,\%$ and $60\,\%$
for most of them. Only two nodes ($150$ and $255$) are insiders one week and linkers the other week.
The linker behavior thus seem rather stable over time, and a one week data set may be enough to
find linkers. To check this hypothesis, we perform simulations of an SIR model using
 data corresponding to one of the two weeks of the study, and use a vaccination strategy targeting the linkers defined
 using data from the other week. Figure \ref{fig:vacc_S} shows that, although this strategy performs less well than the strategy
 using linkers defined from the full two weeks data, it still largely outperforms random vaccination in  reducing the risk of large
 outbreaks and the size of large epidemics. The reduced efficiency obtained when using data from one week to define a strategy
 applied in the other week is reminiscent of the results of \cite{Starnini:2013} showing a limit in the efficiency of vaccination strategies
 due to the fluctuations of contact networks.

\section{Discussion}

In this paper, we analyzed data describing the contacts between individuals in the context of an office building.
Not all individuals participated in the study, which means that the data cover only a part of the real contact network. As the sampled fraction is quite high, and uniform among the departments, we assume that the behavior of the sampled population is representative of the whole.

Under this caveat, we have shown that, in this
situation, many statistical features of contacts are similar to the ones measured in different contexts, with some important specificities.
First, contacts are very sparse, which for instance tends to hinder the spread of infectious diseases, especially rapidly spreading ones.
Second, the  contact network is structured in communities that approximately match the organization in departments of the building.
The connections between departments do not reflect their spatial organization in the building, but rather their roles:
the three scientific departments form a cluster, whereas human resources and logistics are more isolated. The structure is thus very different
from a homogeneous mixing hypothesis, with important consequences on the design of agent-based models of \emph{e.g.} epidemic spreading processes.

The presence of a strong community structure linked to the organization in departments in the contact network has led us
to define three node behaviors, depending on the fraction of links each node
shares within its own department: residents, whose links are mostly internal; wanderers, whose links are mostly external; linkers, whose links are
half internal, half external, and therefore connect departments, acting as bridges between communities \cite{Wasserman:1994,Salathe:2010_PLoS}. Empirically,
the most central nodes of the network turn out to be linkers. As a consequence, targeted vaccination strategies based on the linker
criterion efficiently prevents epidemic outbreaks.
This behavior is also stable enough on the scale of one week for such a strategy to remain efficient if based on fewer data.

The precise identification of linker behavior relies on the knowledge of the contact network. One could thus argue that it requires
almost as much knowledge as the identification of the nodes with highest betweenness centrality.
However, the linker behavior, contrary to the betweenness centrality criterion, may be more easily linked to recognizable human behavior or to
individual attributes in the organizational chart, for instance in relation to professional grade or specific activities. In this case,
one could \emph{a priori} discern which individuals are more susceptible to be linkers and play an important role in the event of an outbreak,
and therefore use such limited information to design an efficient vaccination strategy entailing only a low cost in terms of necessary information
\cite{Smieszek:2013,Chowell:2013}. We final note that the linker behavior might also be identified from limited information in
other social contexts with communities -- schools, hospitals, etc -- and provide an
important ingredient in agent-based models of epidemic spreading phenomena, as such agents provide
crucial gateways between communities. Moreover, in the perspective of modelling and studying epidemic processes in
real, large-scale systems, obtaining a complete data set of contacts between individuals in a large-scale population seems beyond reach, so that this type of
 criterion could be a way to uncover the central elements who could be targeted for outbreak detection and control.

\section{Acknowledgments}

We thank the direction of the InVS for accepting to host this study in their facility, and all the participants to the study.
We acknowledge the SocioPatterns\footnote{\protect\url{http://www.sociopatterns.org/}} collaboration for
providing the sensing platform that was used in the collection of the contact data.
M.G., I.B. and A.B. are partially supported by the French ANR project HarMS-flu (ANR-12-MONU-0018);
C.L.V. and A.B. are partially supported by the EU FET project Multiplex 317532.

\section{Supplementary information}

\subsection{Contact matrices}

\begin{figure*}
  \includegraphics[width=\textwidth]{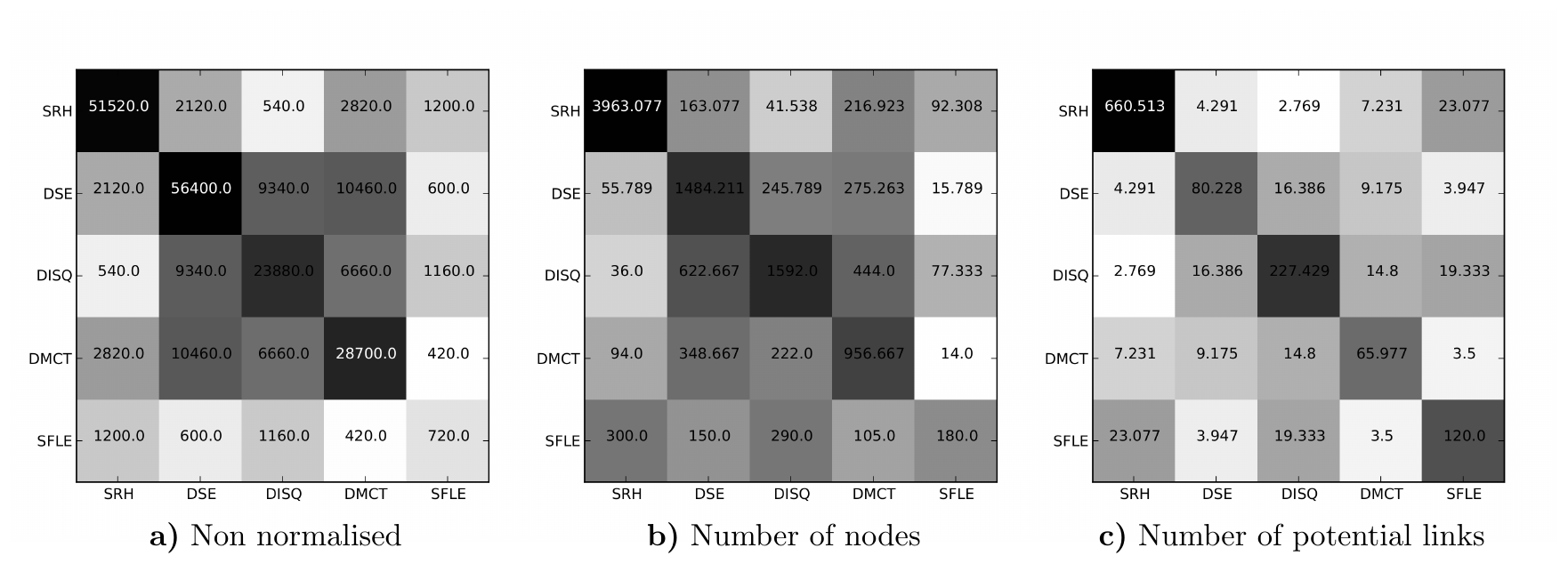}
  \caption{\label{fig:SM:CMN}Original and normalized contact matrices for the whole data set.}
\end{figure*}

Each element of the contact matrix shown in the main text
represents the total time of contact between nodes from two departments.
Since the department populations do not have the same size, it is also of interest to compute normalized contact matrices. We
show in Fig.~\ref{fig:SM:CMN} two different normalizations: in the first one, we divide the elements of each row
by the number of people in the corresponding department. Each element thus represents the mean time each
node from the row department has spent with nodes from the column department (in this case, the matrix is no longer symmetrical).
In the second normalization procedure, we divide each element of the original matrix
by the number of potential links between the two departments. Each element gives then the mean contact duration
between individuals of the two departments.

\begin{figure}
  \includegraphics[width=1.1\textwidth]{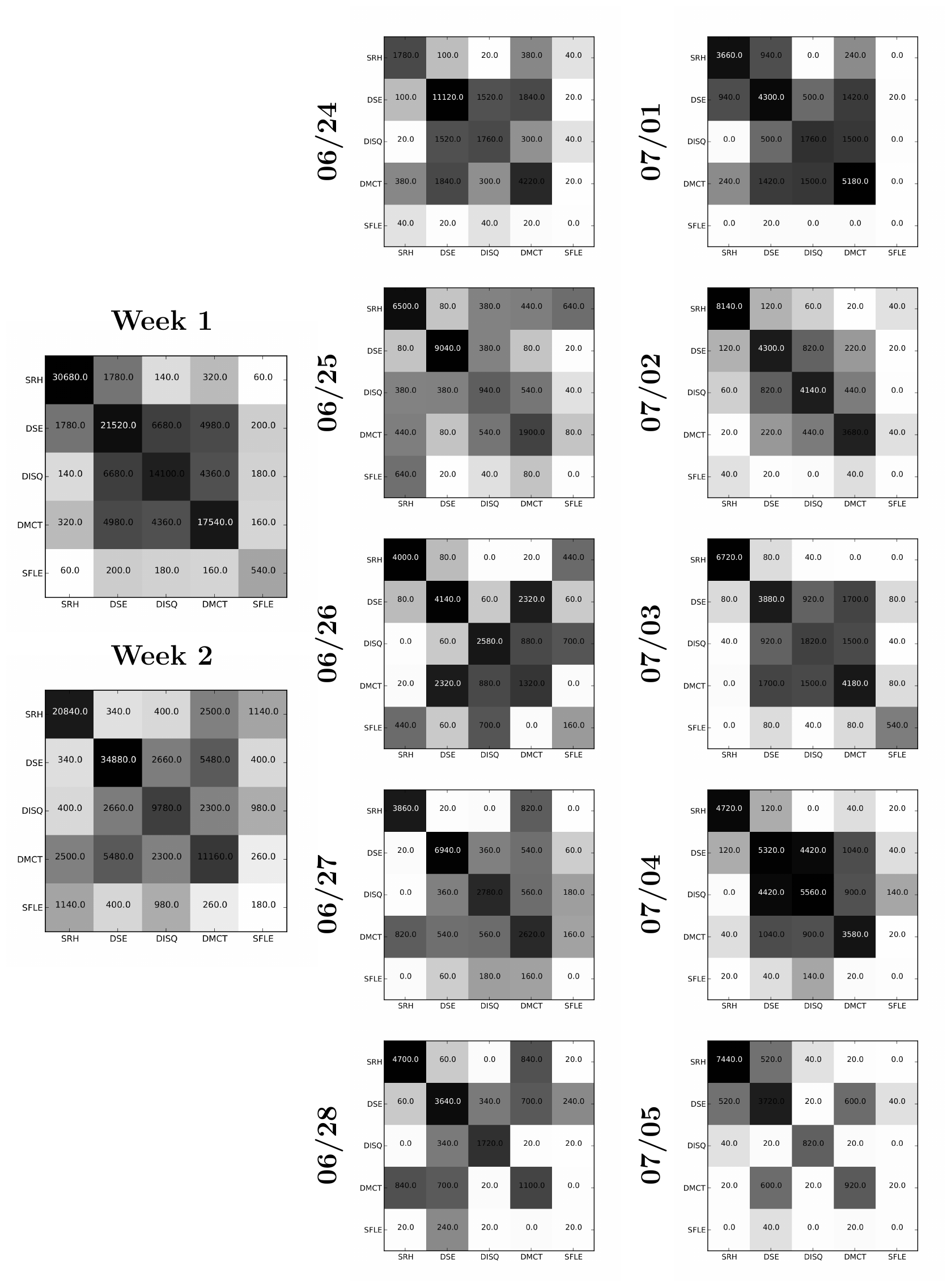}
  \caption{\label{fig:SM:CMT}Weekly and daily contact matrices.}
\end{figure}

For completeness, we therefore also compute the
similarities between daily contact matrices normalized in different ways (Fig.~\ref{fig:SM:CMT} gives the weekly and daily original -- symmetric -- contact matrices).
The values, given in Table \ref{tab:SM:sim_CM}, depend on the specific normalization procedure but remain large.

\begin{table}
  \caption{\label{tab:SM:sim_CM}{\bf Daily contact matrices similarities.} For each day, we compute the cosine similarity between the corresponding contact matrix and the ones of the other days, both with and without diagonals, and for three different normalizations: no normalization, normalization of the rows by the number of nodes in each department, normalization of each element by the number of possible links between the two departments. We list in this table the mean value and the standard deviation of the similarities.}
  \begin{tabular}{ccccccc}
    \hline\hline
    &\multicolumn{2}{c}{No normalization}&\multicolumn{2}{c}{Number of nodes}&\multicolumn{2}{c}{Number of links}\\
    \hline
    Day&full&w/o diagonal&full&w/o diagonal&full&w/o diagonal\\
    \hline
    06/24&0.753 $\pm$ 0.103&0.563 $\pm$ 0.222&0.649 $\pm$ 0.098&0.437 $\pm$ 0.258&0.691 $\pm$ 0.277&0.494 $\pm$ 0.197\\
    06/25&0.843 $\pm$ 0.069&0.481 $\pm$ 0.087&0.748 $\pm$ 0.084&0.297 $\pm$ 0.225&0.707 $\pm$ 0.276&0.301 $\pm$ 0.199\\
    06/26&0.837 $\pm$ 0.046&0.400 $\pm$ 0.246&0.637 $\pm$ 0.072&0.207 $\pm$ 0.157&0.421 $\pm$ 0.220&0.329 $\pm$ 0.151\\
    06/27&0.870 $\pm$ 0.052&0.534 $\pm$ 0.108&0.819 $\pm$ 0.051&0.380 $\pm$ 0.091&0.762 $\pm$ 0.297&0.370 $\pm$ 0.184\\
    06/28&0.871 $\pm$ 0.075&0.426 $\pm$ 0.134&0.804 $\pm$ 0.110&0.341 $\pm$ 0.220&0.798 $\pm$ 0.171&0.324 $\pm$ 0.086\\
    07/01&0.821 $\pm$ 0.058&0.592 $\pm$ 0.152&0.788 $\pm$ 0.083&0.437 $\pm$ 0.234&0.763 $\pm$ 0.302&0.431 $\pm$ 0.199\\
    07/02&0.850 $\pm$ 0.087&0.579 $\pm$ 0.180&0.827 $\pm$ 0.110&0.463 $\pm$ 0.231&0.772 $\pm$ 0.305&0.510 $\pm$ 0.185\\
    07/03&0.858 $\pm$ 0.072&0.488 $\pm$ 0.262&0.770 $\pm$ 0.096&0.376 $\pm$ 0.225&0.273 $\pm$ 0.275&0.418 $\pm$ 0.224\\
    07/04&0.767 $\pm$ 0.058&0.317 $\pm$ 0.131&0.704 $\pm$ 0.089&0.219 $\pm$ 0.111&0.679 $\pm$ 0.277&0.255 $\pm$ 0.138\\
    07/05&0.795 $\pm$ 0.123&0.398 $\pm$ 0.199&0.762 $\pm$ 0.148&0.271 $\pm$ 0.177&0.725 $\pm$ 0.294&0.387 $\pm$ 0.176\\
    \hline\hline
  \end{tabular}
\end{table}

\begin{figure}
  \includegraphics[width=1.1\textwidth]{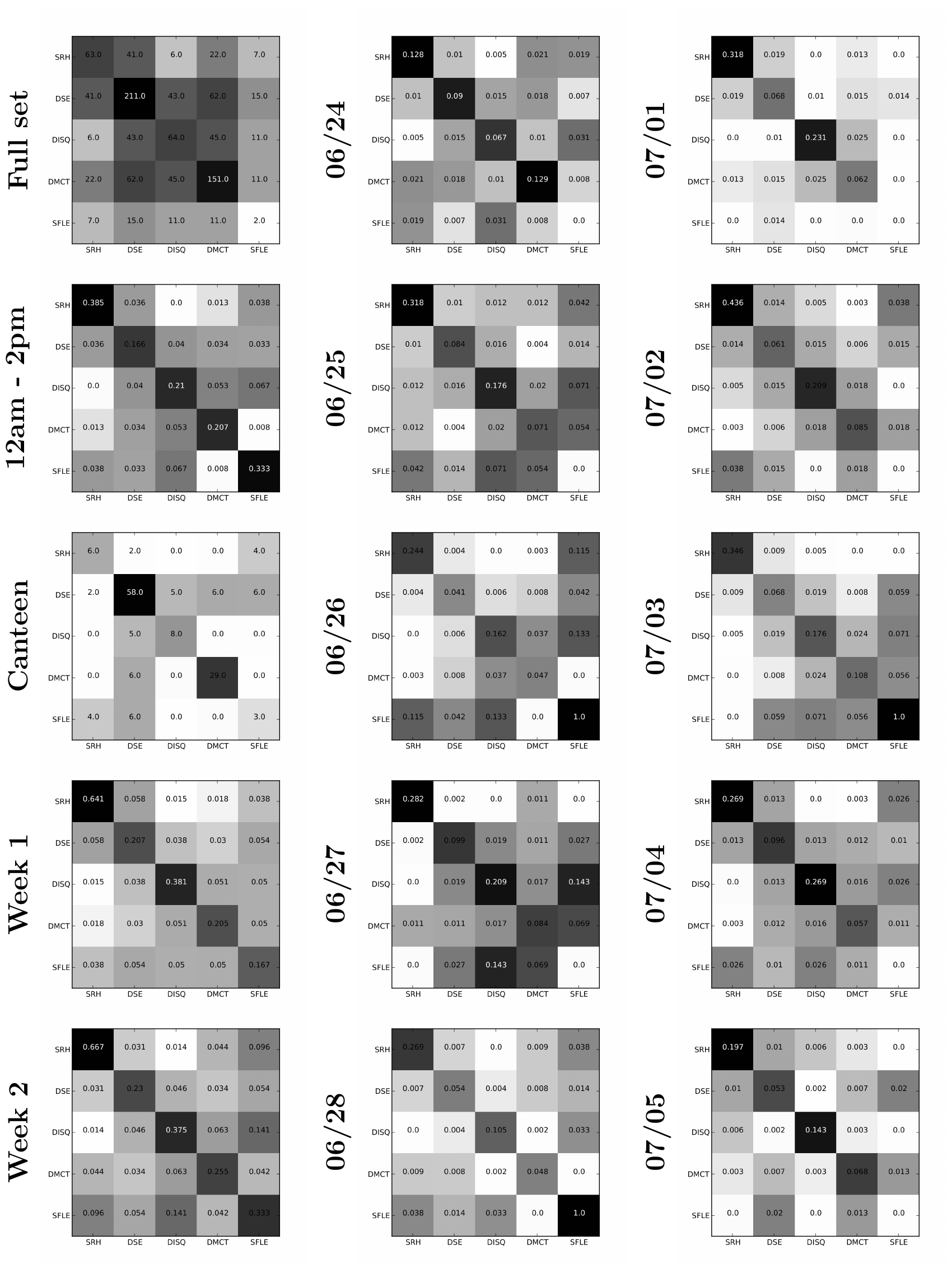}
  \caption{\label{fig:SM:CML}Link density contact matrices.}
\end{figure}

If the ratio of the number of links over the total number of possible links is considered instead of the contact durations, contact matrices represent the connectivity of the contact network (Fig.~\ref{fig:SM:CML}).

\begin{figure}
  \includegraphics[width=1.1\textwidth]{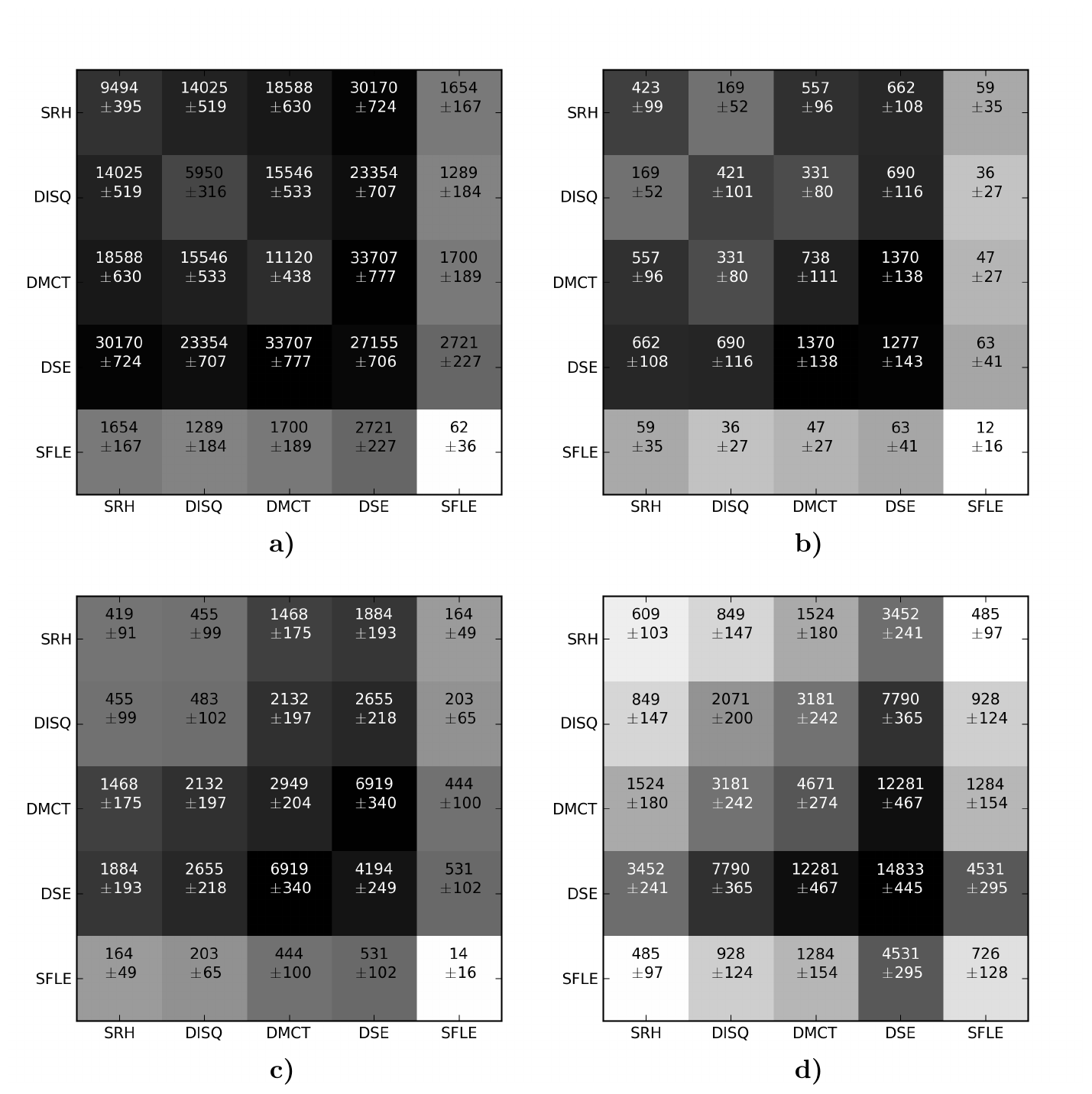}
  \caption{\label{fig:SM:CM_nm}Contact matrices: null model with constant contact probability and empirical presence timelines, for $\Delta = 30$\,min, averaged over 100 different realizations.
Each matrix element (at row X and column Y) gives
  the total time of contact (mean $\pm$ s.e.m.) between individuals from  departments X and Y
  during the two weeks of the study, in different locations, according to the null model.
  {\bf a)} Entire building. {\bf b)} Conference room. {\bf c)} Cafeteria, restricted to the interval between 12am and 2pm for each day.
  {\bf d)} Canteen. This place is in a different building and was not taken into account in a).}
\end{figure}

In order to test the effect of spatial organization on the shape of the contact matrix,
we build a null model where only the timeline of presence of each node is taken into account, and interactions are assumed to take place at random
between individuals who are in the same location.
The timelines of presence are built from the empirical contact data: a node is present at a given location at a time $t$ if it takes part in a contact recorded here at this time. A node that is present at $t$ is moreover assumed to be present during the interval $[t-\Delta,t+\Delta]$. At each time $t$, all nodes that are present in a given location have a constant probability to be in contact. The probability is chosen such that the total cumulative duration of contacts is equal to its empirical value. The contact matrix obtained, shown in
Fig. \ref{fig:SM:CM_nm}, is significantly different from the empirical one. This
indicates that the empirical contact matrix
structure is not explained by random encounters of individuals with different presence timelines.

\clearpage
\newpage

\subsection{Effect of data representation on epidemic spreading.}

The collected contact data consist in a temporal network at a very high temporal resolution. These data can be
aggregated and represented in different ways, both along the temporal and the organizational dimensions, in order e.g.
to build models for the spread of epidemics in the population.  As discussed e.g. in
\cite{rep_1,rep_2,rep_3,rep_4,anna,question}, the level of detail of the data representation that is taken into account
in the model can influence the outcome of the simulations. We consider this issue with the data at hand, by using
five different representations for the contact dynamics, from highly detailed to simplistic, and by using each representation
as the support of an SIR model for epidemic spread \cite{anna}:
\begin{itemize}
\item {\bf Full data.} We use the temporal network built from the empirical data at the highest temporal resolution (20\,s).

\item {\bf Heterogeneous static network.} We use the contact network aggregated over the whole data collection period: in this network,
nodes representing individuals who have been in contact at least once are connected by a link whose weight is given by the total
contact time of these individuals, normalized by the total duration of the data set.

\item {\bf Global contact matrix.} We consider that all nodes are connected to each other, and that the weight of a link connecting
two nodes depends only on their respective departments: it is given by the average contact time of all pairs of individuals belonging to
these departments. In other words, the total contact time between each pair of departments is equally redistributed among all
pairs of individuals of these departments.

\item {\bf Daily contact matrices.} We consider a contact matrix representation, using for each day the corresponding daily contact
matrix to compute the weights of the links between individuals.

\item {\bf Homogeneous mixing.} We consider a fully connected contact network with homogeneous weights, computed as the average
contact time between any two individuals (independently of their department).

\end{itemize}
In each representation, we moreover take into account inactivity periods (nights and weekends) by assuming that all nodes are isolated
during these periods.

\begin{figure}
  \includegraphics[width=\textwidth]{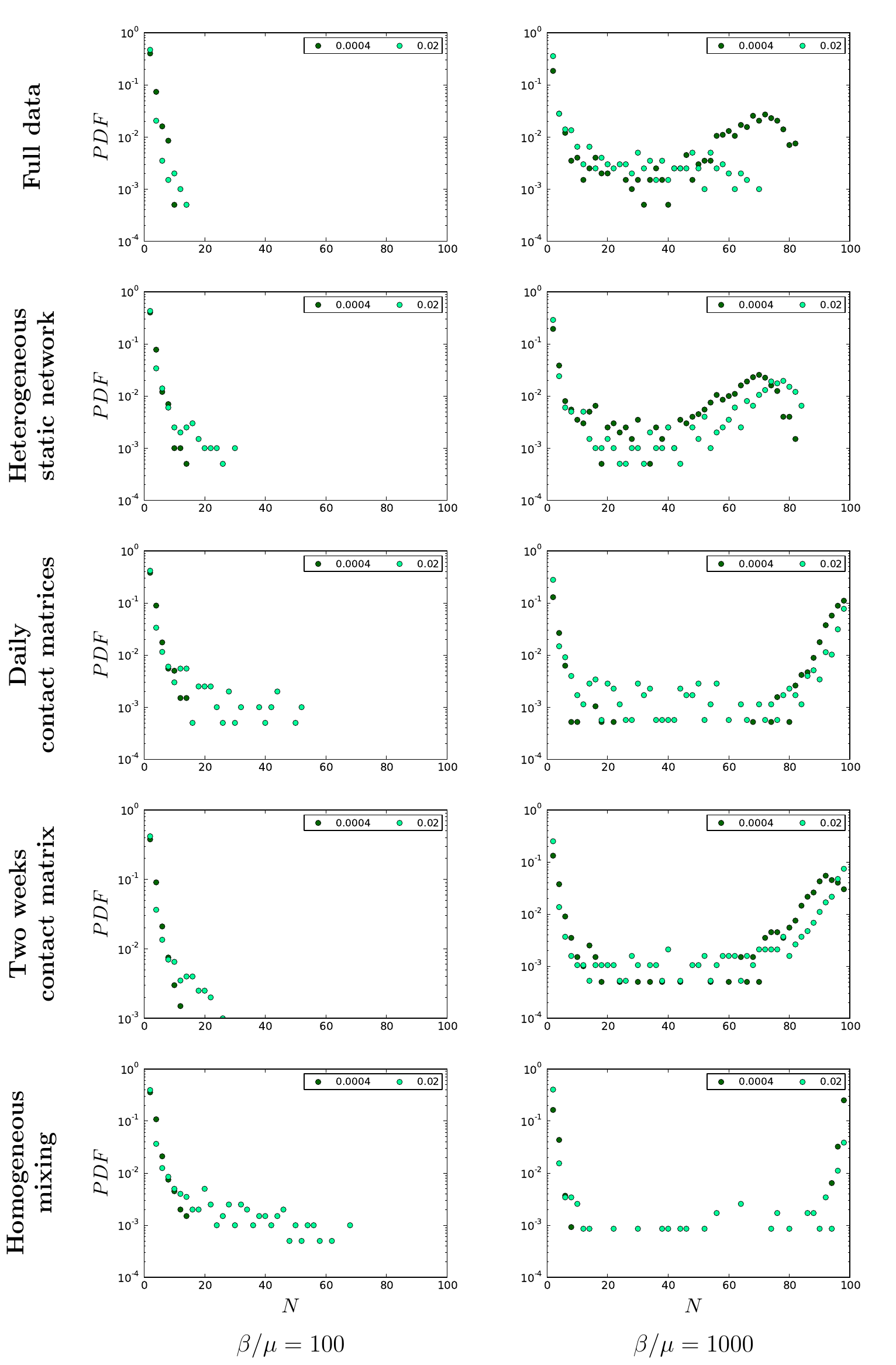}
  \caption[S]{\label{fig:SM:epidemics}{\bf Distributions of the size $N$ of epidemics.} Simulations are done for different values of the infection rate $\beta$, the recovering rate $\mu$ being fixed by the constant $\beta/\mu$ ratio. For each value of $\beta$, statistics are computed from 1000 simulations.}
\end{figure}

The results of the numerical simulations  of an SIR model are shown in Fig.~\ref{fig:SM:epidemics} for two values of $\beta/\mu$.
For $\beta/\mu = 100$, no matter which representation is used, most of the epidemics do not reach a large fraction of the population.
The tails of the distribution of epidemic sizes however become broader when using contact matrices or a homogeneous mixing assumption,
as the sparsity of the contacts is then not correctly considered \cite{anna}.

This effect is seen most clearly for $\beta/\mu = 1000$.
With complete contact information, i.e., if the spread is simulated on the time-resolved contact network,
the distribution depends on the value of $\beta$, as discussed in the main text. For small
$\beta$ (slow epidemics), a second mode develops at large values of the epidemic size. For faster epidemics, this second mode
is suppressed: the recovery time becomes small enough for the temporal contact patterns
to have an impact on the spread. Much less infection paths are available during the infectious period of each node,
and thus the epidemics does not spread as much
as when $\beta$ and $\mu$ are small (in which case a node remains infectious for a longer time implying that it has more contacts during its
infectious period and the disease has more occasions to spread).
When static representations are used, and in particular when using contact matrices or a homogeneous mixing hypothesis, the
second mode is strongly overestimated. The second mode moreover is not suppressed when $\beta$ increases, and even
shows the opposite tendency with respect to the time-resolved
network: as the epidemic spreads faster, it unfolds over a smaller number of inactivity periods (nights and week-ends) and
therefore tends to reach larger sizes; on the other hand, the spread during the days does not depend on $\beta$, at fixed $\beta/\mu$.

Overall, these results are similar to the ones obtained in \cite{anna} in a different context and show the importance of using a data representation that
includes enough information on the sparsity and heterogeneity of contact networks.

\end{document}